\documentclass{elsart}
\journal{New Astronomy}
\usepackage{natbib}

\usepackage{graphicx}
\usepackage{epsfig}

\usepackage{amssymb}

\begin{document}

\begin{frontmatter}



%\title{Quadrupole Sciences with CMB Polarization}
 \title{CMB-induced Cluster Polarization as a
Cosmological Probe}


\author{Daniel Baumann$^1$ and Asantha Cooray$^2$}
\address{$^1$DAMTP, Centre for Mathematical Sciences, University of Cambridge, Wilberforce Road, Cambridge CB3 0WA, United Kingdom\\
$^2$Theoretical Astrophysics, California Institute of Technology, Pasadena CA 91125\\
E-mail: db275@cam.ac.uk, asante@caltech.edu}

\begin{abstract}
% Text of abstract
Scattering of the temperature anisotropy quadrupole by
free electrons in galaxy clusters leads to a secondary
polarization signal in the cosmic microwave background (CMB) fluctuations.
At low redshifts, the temperature quadrupole contains
a significant contribution from the integrated Sachs-Wolfe (ISW) effect 
associated with the growth of density fluctuations.
Using polarization data from a sample of clusters over a wide range in
redshift, one can statistically establish the presence of the ISW effect
and determine its redshift evolution. 
Given the strong dependence of the ISW effect on the background
cosmology, cluster polarization can eventually be used as a powerful
probe of dark energy. As a further application, we also discuss how it
might be used to understand the
potential lack of power on large scales.
\end{abstract}


\end{frontmatter}

% main text
\section{Introduction}
Polarization of cosmic microwave background (CMB) anisotropies is only
generated when the CMB photons scatter off free electrons. The
polarization therefore traces the ionization history of the
Universe. The Universe was fully ionized at early times, before last
scattering occurred at a redshift of about $z_{rec} \approx 1100$,
after which the Universe became neutral and radiation and matter
decoupled. At that time the {\it primary} polarization of the
microwave background radiation was generated. Since polarization is a strictly
causal process this signal is expected to peak at the horizon scale at
recombination. Causality doesn't allow a polarization signal on larger
scales. Such a signal, however, has recently been measured by the
Wilkinson Microwave Anisotropy Probe (WMAP) \cite{kogut}. This
large-scale {\it secondary} polarization is interpreted as a
signature of a late time reionization of the Universe at a redshift of
$z \sim 10-20$. Reionization leads to free electrons in galaxy
clusters and hence small-scale secondary polarization. The
contribution of {\it unresolved} clusters to the polarization power
spectrum has been calculated in \cite{BauCooKam}.
Galaxy clusters produce equal E- and B-mode polarization, but at a level that
is orders of magnitude below the primary polarization and clusters are
therefore unlikely to be a source of confusion for future CMB
experiments.\\ 
In \cite{CooBau} we revisited the problem of measuring a
CMB-induced polarization signal towards {\it resolved} galaxy
clusters. This was previously studied by e.g. \cite{SunZel80},
\cite{SazSun}, \cite{Challinor}, and \cite{Audit}. 

Linear polarization of the cosmic microwave background is generated
through rescattering of the temperature quadrupole.
In a cosmological model with dark energy the quadrupole evolves
between the last scattering surface ($z=1100$) and us ($z=0$) due the
integrated Sachs-Wolfe (ISW) effect. The quadrupole-induced
polarization signal therefore probes dark energy through the ISW
effect. Kamionkowski and 
Loeb~\cite{KamLoeb} have considered the possibility of using multiple such measurements to reduce the cosmic variance 
uncertainty in the CMB temperature quadrupole.
The connection between properties of dark energy, the
quadrupole at the cluster redshift and polarization in the direction
of galaxy clusters is presented in this contribution 
to these proceedings and its applications are discussed.

\section{CMB-induced Polarization towards Clusters}
\label{sec:pol}

CMB polarization towards clusters is generated when the incident radiation has a nonzero quadrupole moment. The two dominant origins for this quadrupole moment are: (a) a projection of the {\it primordial} CMB quadrupole to the cluster location, and (b) a local {\it kinematic} quadrupole from cluster peculiar motion.
Towards a sufficiently large sample of galaxy clusters, we can write the total
rms degree of polarization as $P_{\rm Total}^2 = P_{\rm Prim}^2 + P_{\rm Kin}^2$,~\cite{SazSun}, where
\begin{equation}\label{e:prim}
P_{\rm Prim} \propto \langle \tau \rangle \,Q^{\rm rms}(z) \propto \sqrt{C_2(z)} \, ,
\end{equation}
\begin{equation}
P_{\rm Kin} \propto g(x)\, \langle \tau \rangle \,\langle
\beta_t^2\rangle \, . 
\end{equation}
$\tau =\sigma_T\int dy  n_e(y) $ is the scattering optical depth of each cluster. Since we are averaging over large samples of clusters,
we consider the sample-averaged optical depth, $\langle \tau \rangle$.
$\beta_t=v_t/c$ gives the transverse component of the cluster velocity
and $g(x)=(x/2)\coth(x/2)$, with $x \equiv h\nu/k_B T_{\rm CMB}$, is the frequency dependence of the kinematic effect.
With the optical depth in individual clusters determined by other
methods, such as the Sunyaev-Zel´dovich (SZ, \cite{SunZel80}) effects,
one can invert the measured polarization, equation~(\ref{e:prim}), to
obtain the rms CMB-quadrupole, $Q^{\rm rms} (z) = (5 C_2/4\pi)^{1/2}$, at the cluster redshift.

The primordial CMB-quadrupole is dominated by two effects.
The Sachs-Wolfe (SW, \cite{SacWol67}) effect arises as a combination of gravitational
redshift and time-dilation effects and can be viewed as a direct
projection of the conditions at last scattering with no evolution
after that time:
\begin{equation}
\left(\frac{\Delta T}{T}\right)^{{\rm SW}} = {\Phi \over 3}_{\tau=\tau_{\rm ls}}\, .
\end{equation}

The integrated Sachs-Wolfe (ISW) contribution arises along the photon path from the time of last scattering to today, as
the CMB photons pass through a time-varying potential: 
\begin{equation}
\left(\frac{\Delta T}{T}\right)^{{\rm ISW}} =  2 \int_{\tau_{\rm ls}}^{\tau_0}
\dot{\Phi} \, d\tau\, . 
\end{equation}
Effectively, the photon receives a shift in energy because the potential it falls into is different from the potential it must climb out of.
The ISW effect is absent in a matter-dominated, critical-density
universe (Einstein-de Sitter). In a universe with dark energy ($w
\equiv p/\rho < 0$) or a cosmological constant, $\Lambda$, ($w=-1$) the ISW effect leads
to an increase in power on large scales.
The expected redshift evolution of the quadrupole, $C_{l=2}(z) = C_{l=2}^{{\rm SW}}(z)+C_{l=2}^{{\rm ISW}}(z)$, is hence characterized by a rise at low redshifts ($z<1$), the time at which the universe becomes $\Lambda$-dominated.

The origin of the kinematic effect is understood as follows. Consider electrons moving with peculiar velocity, $\beta=v/c$, relative to the rest frame defined by the CMB.
The Doppler-shifted spectral intensity of the CMB in the mean electron rest frame is 
\begin{equation}
I_{\nu} = C \frac{x^3}{e^{x \gamma(1+\beta \mu)}-1}\, ,
\end{equation}
where $x \equiv h \nu/ k_B T_{\rm CMB}$, $\gamma = (1-\beta^2)^{-1/2}$ and
$\mu$ is the cosine of the angle between
the cluster velocity and the direction of the incident CMB photon.
When expanded in terms of Legendre polynomials, the intensity distribution is
\begin{equation}
I_{\nu} = C \frac{x^3}{e^x-1} \left[ I_0 + I_1 \mu +
\frac{e^x(e^x+1)}{2(e^x-1)^2}x^2 \beta^2 \left(\mu^2-\frac{1}{3}\right) + \dots
\right] \, ,
\end{equation}
which contains the necessary quadrupole under which scattering generates
polarization. 

\begin{figure}[!t]
\centerline{\psfig{file=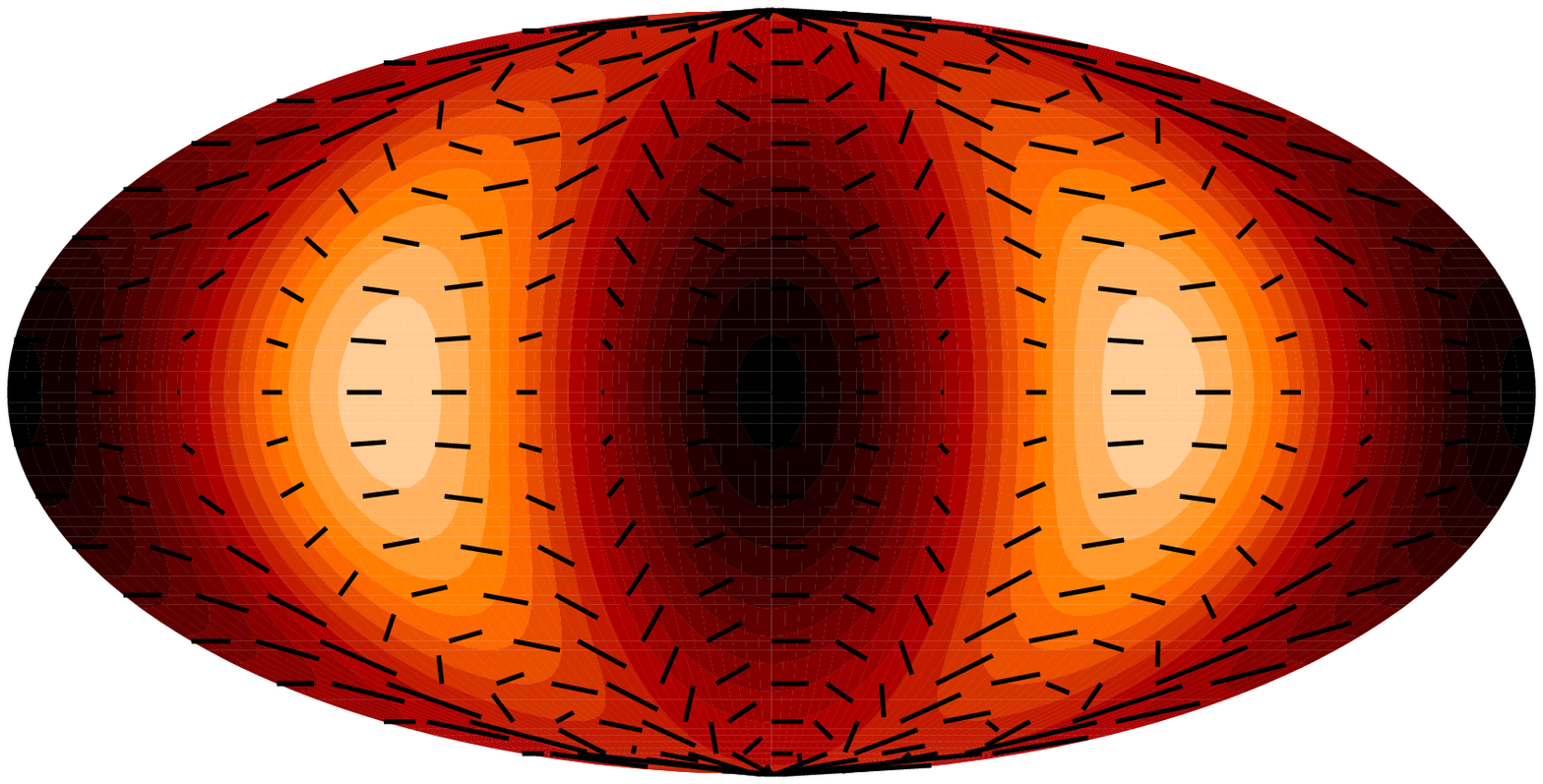,width=2.7in}\psfig{file=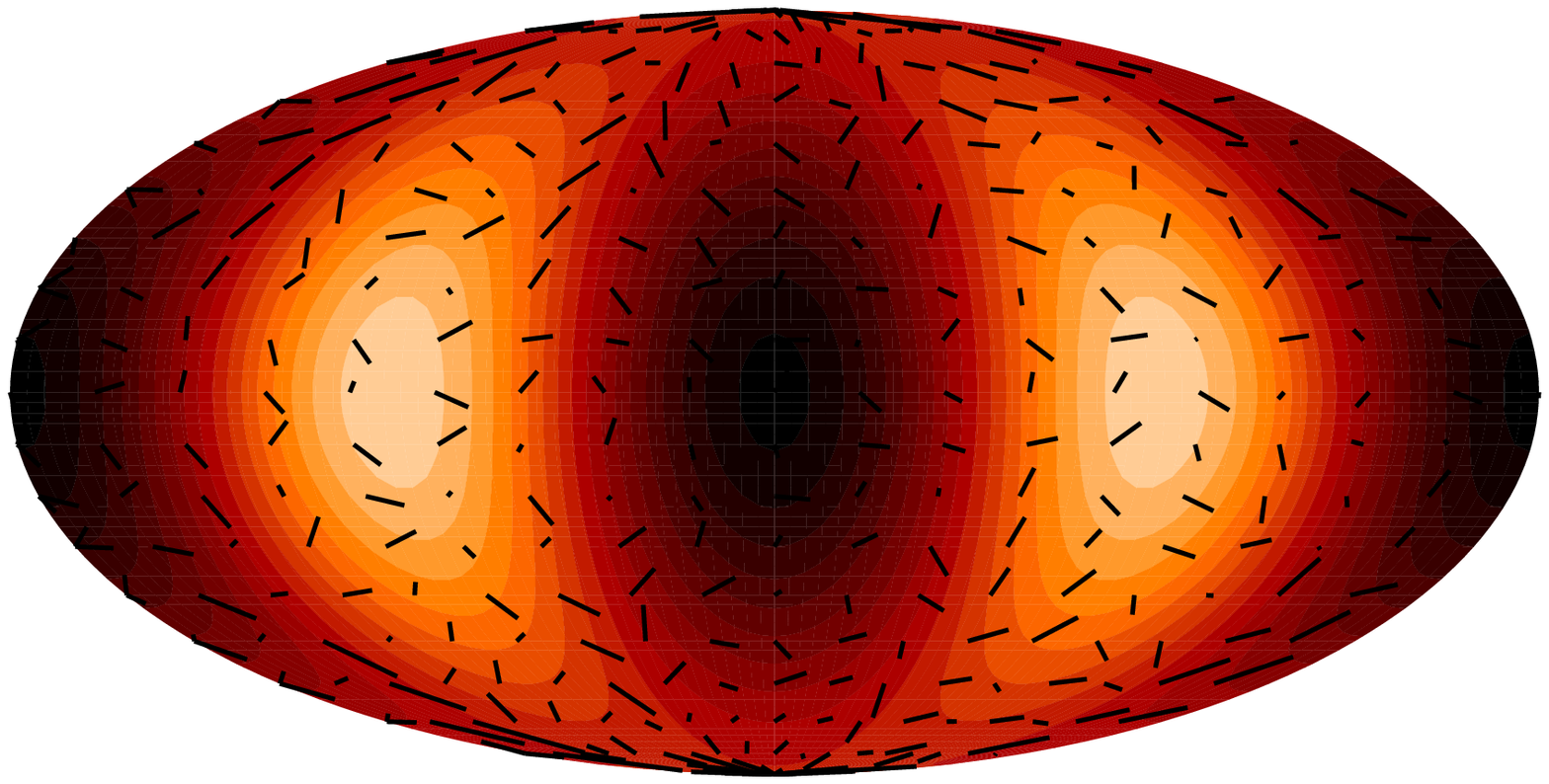,width=2.7in}}
\caption{CMB polarization due to galaxy clusters.
The polarization vectors give a representation of the expected
polarization from a cluster at the corresponding location in the sky.
The scale is such that the maximum length of a line corresponds to a
polarization of 4.9$\tau$ $\mu$K.
The background color represents the temperature quadrupole. The left
map shows
the resulting polarization contribution due to scattering of the
primordial  temperature quadrupole alone.
The map for the total polarization contribution (primordial
and kinematic) is indistinguishable from this map. 
To make the kinematic polarization visible
we arbitrarily increase its amplitude by a factor of 100 (right map).
This ``scale factor'' means that the primordial polarization 
dominates the total contribution even at high frequencies where the
kinematic quadrupole is boosted due to its spectral dependence.}
\label{fig:prim}
\end{figure}

Figure~\ref{fig:prim} illustrates the
primordial and kinematic quadrupole contributions using all-sky
maps of the expected polarization.
In these plots, each polarization vector should be considered as a representation of the polarization towards a
cluster at that location. The polarization pattern created by scattering of the primordial quadrupole is uniform and traces the underlying
temperature quadrupole distribution.
For the kinematic quadrupole we assume, for illustrative purposes,
a transverse velocity field with $\langle \beta_t \rangle \sim
10^{-3}$ corresponding to a velocity of 300 km sec$^{-1}$.
The polarization contribution due to
the kinematic quadrupole, however, is random due to the fact that
transverse velocities are uncorrelated\footnote{The
correlation length of the velocity field ($\sim$ 60 Mpc) correlates
velocities within regions of 1$^{\circ}$ when projected to a redshift of
order unity.}. 
This explains the randomisation of the all-sky polarization map when a
significant kinematic contribution is included.
It should be noted, however, that the kinematic effect has been scaled by a factor of $100$ to make it visible in Fig.~\ref{fig:prim}.
Therefore, as shown in Fig.~\ref{fig:prim},  the primordial polarization  dominates the total contribution even at high frequencies where
the kinematic quadrupole is increased due to its spectral dependence.
Also, the spectral dependence of the kinematic quadrupole contribution,
$g(x)$, gives a potential method to separate the two polarization effects
\cite{Dod97}. 
This is similar to component separation suggestions in
the literature as applied to temperature observations, such as the
separation of the thermal SZ-effect from dominant primordial
fluctuations \cite{Cooetal00}.

\section{Aspects related to the Primordial CMB Quadrupole}
The basic idea is to use galaxy clusters as tracers of the local temperature quadrupole and statistically detect its rms value.
Imagine a future experiment measuring the polarization towards resolved clusters. For each cluster we also measure its redshift and the optical depth through the thermal SZ effect. We do this for a large sample of clusters, bin the resulting data into redshift intervals and average over all sky. If the bin size can be chosen small enough this will allow us to reconstruct the rms temperature quadrupole as a function of redshift. Multi-frequency observations will allow to separate the primordial quadrupole from the contaminant kinematic contribution with only a factor of $\sim 2$ enhancement in the instrumental noise \cite{CooBau}.

For the best-fit $\Lambda$CDM model with $70\%$ dark energy, the
quadrupole leads to a maximum
primordial polarization of $P_{{\rm Prim}} \sim 4.9\tau$ $\mu$K. Since
the kinematic polarization scales as $P_{{\rm Kin}} = 0.27 g(x)
(\beta_t/0.001)^2 \tau$ $\mu$K, we expect the CMB-induced signal to be
dominant (as illustrated in Fig.~\ref{fig:prim}). Factors that could make the primordial and the kinematic polarization more comparable are the frequency boost of the kinematic effect, a more optimistic estimate of $\langle \beta_t\rangle$ and a low value of the CMB-quadrupole.
However, even if the signals were of comparable magnitude the random
orientations and characteristic frequency dependence of the kinematic
polarization would allow a reliable extraction of the primordial
CMB-quadrupole.
\begin{figure}[!t]
\centerline{\psfig{file=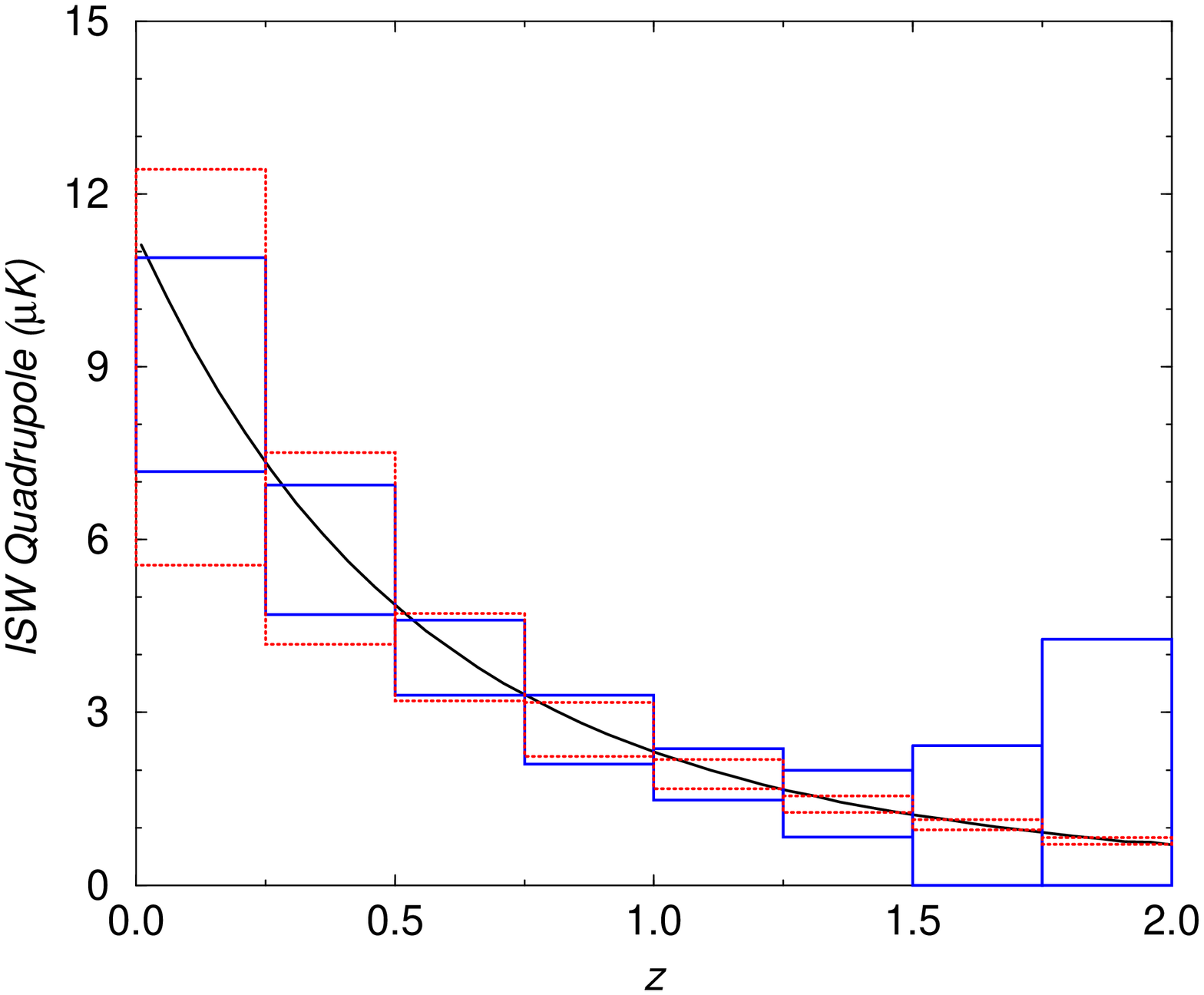, width=2.2in, angle=-90}
\psfig{file=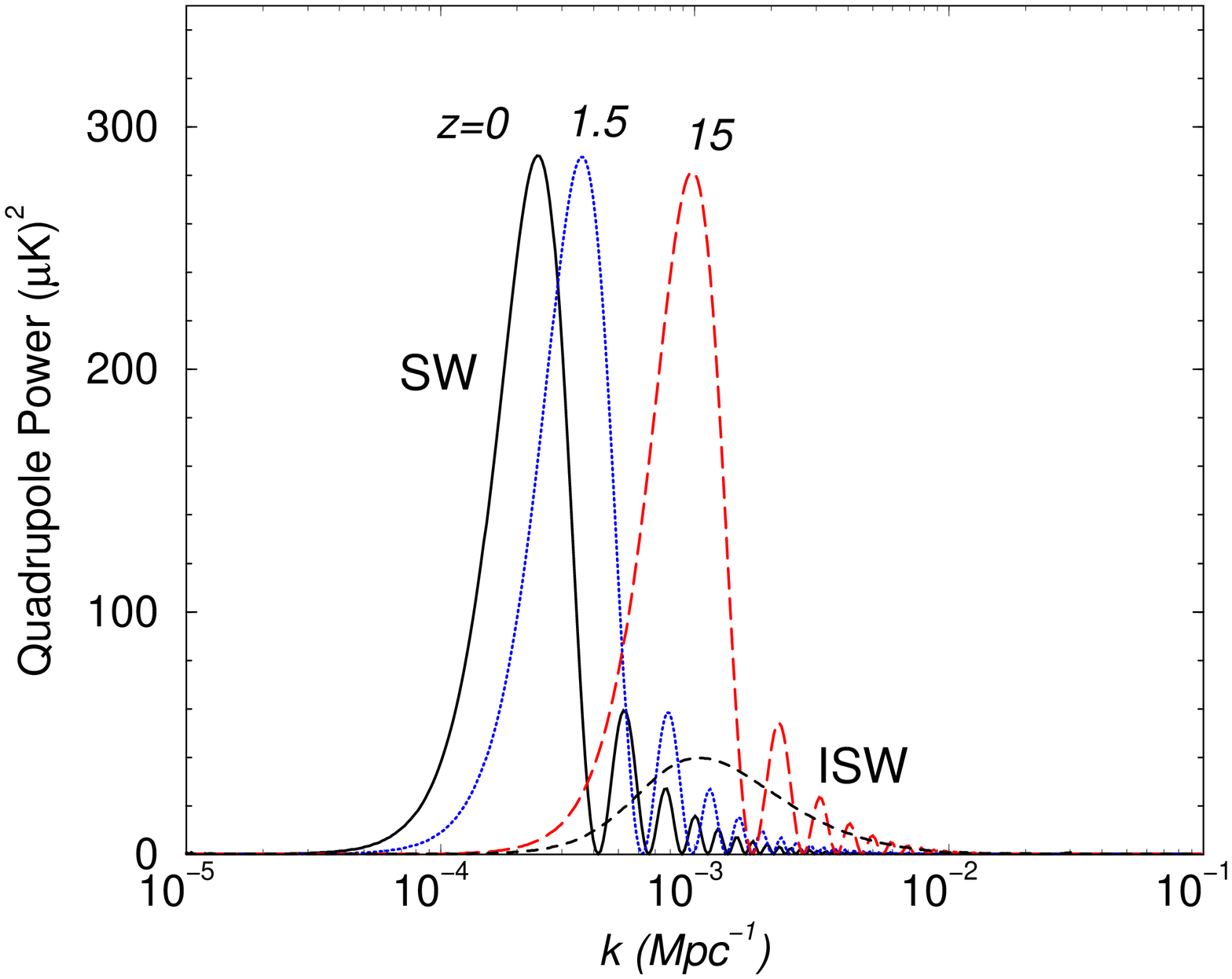, width=2.2in, angle=-90}}
\caption{{\it Left:} Statistical reconstruction of the ISW contribution to the
redshift evolution of the temperature quadrupole \cite{CooHutBau}. 
The solid error bars assume a reconstruction with clusters down to
10$^{14}$ M$_{\rm Sun}$ in an area of 10$^4$ deg$^2$ with an instrumental
noise of 0.1 $\mu$K. The dotted lines show the cosmic-variance for an
all-sky reconstruction computed from the number of independent volumes
sampled by clusters at each redshift bin~\cite{KamLoeb}.
{\it Right:} The projected power in the temperature quadrupole as a function of the
wave number. The rms quadrupole $Q^{\rm rms} (z)$
is given by an integral of these functions. We consider three redshifts ($z=0$, $1.5$ and $15$). 
Since the quadrupole at low redshift is similar to the one we observe, it is potentially affected by any suppression of power at large scales. 
The low redshift cluster polarization may therefore be exploited as a check on the quadrupole measurement by WMAP.}
\label{fig:qrms}
\end{figure}

In \cite{CooBau}, we assessed the potential detectability of the ISW
signature with cluster polarization (Figure~\ref{fig:qrms}), while
in \cite{CooHutBau}, we considered the measurement of cosmological parameters related to the dark energy
using cluster polarization data from, say, the planned CMBpol mission. 
As discussed there, the measurement of dark energy properties is aided by the fact that
one probes the redshift evolution of the ISW contribution through the growth rate of the gravitational potential.
The latter provides the most sensitive probe of dark energy when
compared to all other cosmological probes considered so far. This
comes from the fact that the growth rate of the gravitational potential is directly proportional to
the dark energy equation of state, while quantities such as the
distance or the growth factor involve, at least, one integral of this
quantity.

To highlight the difference between the quadrupole we observe today
and the local quadrupole inducing the cluster polarization, we also show,
in Fig.~\ref{fig:qrms}, the power of the projected quadrupole, as a function of
wavenumber $k$, at several redshifts. Note that the 
primordial SW contribution to the projected quadrupole contains mostly
information from large scales.
There are now indications, based on the WMAP data, that there is a significant suppression of
power on scales corresponding to $k < 5 \times 10^{-4}$ Mpc$^{-1}$
\cite{Contaldi:2003zv}. At low redshifts, the projected quadrupole related
to cluster polarization
would also be suppressed, since it is strongly correlated with the one we observe. As one moves to
higher redshift, however, this suppression will become less
significant, since the projection effects move the dominant
contribution to smaller scales (Figure~\ref{fig:qrms}).
Thus, at the expected reionization redshift of order 20, consistent with WMAP's optical depth of $\tau=0.17$,
the quadrupole that generated the large angular scale CMB polarization
is {\it not} affected by the suppression of power one sees in the quadrupole
observed today. While the primordial quadrupole today and at low redshifts may be affected by some unknown physics, 
the ISW contribution to the quadrupole is expected to remain
unaffected, since its contribution arises from smaller scales (Figure~\ref{fig:qrms}). 
The total polarization signal, however, will be smaller and if this happens to be the case towards low redshift clusters,
one can be certain that the observed suppression of power at largest angular scales is not related to
potential systematics in the WMAP data, such as due to the galaxy cut.

\section{Discussion}
\label{sec:discussion}
Establishing the ISW effect and reconstructing its redshift evolution poses a significant experimental challenge.
The advent of polarization sensitive bolometers, however, suggests
that a reliable reconstruction of the ISW signature is within reach
over the next decade. Given the strong dependence of the ISW effect on the background
cosmology, cluster polarization can eventually be used as a powerful probe of
the dark energy. A detailed analysis of the potential of this method
for extracting information on the dark energy equation of state,
$w(z)$, is presented in \cite{CooHutBau}.

Finally, we remarked on a further application of this method related to the recent WMAP results.
The $\Lambda$CDM concordance model fits the WMAP data perfectly
on small and intermediate scales. However, there are indications that
the data is inconsistent with the simplest such models on the largest
angular scales. The lack of power in CMB temperature fluctuations on large angular
scales ($\ell<$ few) is intriguing. \cite{Spergel:2003cb}
showed that the probability of obtaining the low-$\ell$ data given the best-fit $\Lambda$CDM theory is
exceedingly small. Future cluster polarization studies may be used to
independently confirm this result, and to exactly determine the physical scale at which this suppression occurs.

The authors thank Dragan Huterer and Marc Kamionkowski for collaborative work and stimulating discussions.

\end{document}